\documentclass[%
superscriptaddress,
onecolumn,
 amsmath,amssymb,
 aps,
11pt,
tightenlines]{revtex4-2}

\usepackage{graphicx}
\usepackage{dcolumn}
\usepackage{bm}
\usepackage{dsfont}
\usepackage{physics}
\usepackage{enumitem}
\usepackage{hyperref}
\usepackage{xcolor}

\begin{document}

\preprint{APS/123-QED}

\title{Quantum resource redistribution drives spectral splits in dense neutrino gases}

\author{Michael Hite}
 \email{michaelhite@arizona.edu}
 \affiliation{Department of Physics, University of Arizona, Tucson, AZ, USA 85721}
\author{Pooja Siwach}%
 \email{siwach@arizona.edu}
\affiliation{Department of Physics, University of Arizona, Tucson, AZ, USA 85721}
\affiliation{Facility for Rare Isotope Beams, Michigan State University, East Lansing, Michigan 48824, USA}

\date{\today}

\begin{abstract}
Whether a quantum many-body system can be efficiently simulated hinges not only on its size but also on how quantum resources are organized within it. We characterize the quantum resource landscape of collective neutrino oscillations using entanglement entropy, non-local magic, and matrix product state bond dimension. Using tensor network simulations of neutrinos in the two-flavor sector, we demonstrate that spectral splits--sharp energy-dependent flavor swaps--emerge precisely where entanglement entropy is maximized and non-local magic is minimized locally. This anticorrelation reveals that spectral splits arise not from generic resource growth but from a structured redistribution among flavor modes. The resource dynamics trace constrained arcs in the entanglement-magic phase space, bounded by entanglement spectrum normalization. These findings establish a direct, quantitative link between quantum resources governing computational complexity and astrophysical observables, informing the design of tensor network and quantum circuit simulations of dense neutrino environments.
\end{abstract}

\maketitle


\section{Introduction}
Neutrinos are the primary agents of energy transport in core-collapse supernovae~\cite{janka:arnp2016,akaho:prl2026}. In the dense regions above the neutrinosphere, neutrino-neutrino forward scattering gives rise to collective flavor oscillations~\cite{Duan:2010bg}—a many-body phenomenon whose full quantum features, including entanglement and multi-particle correlations, are not captured by mean-field treatments~\cite{Balantekin:2023qvm}. A striking manifestation of these collective dynamics is the emergence of spectral splits: sharp, energy-dependent swaps of flavor content in the neutrino spectra \cite{Patwardhan:2021rej,siwach:2023prd3,Duan:jpg2009}. Restricting to two neutrino flavors, the resulting Hamiltonian can be mapped onto an all-to-all spin model amenable to both classical tensor network~\cite{Cervia:2022pro} and quantum simulation methods~\cite{Hall:2021rbv,siwach:prd2023}. In the three-flavor framework, neutrinos naturally map to qutrit degrees of freedom~\cite{lacroix:prd2026,heimsoth:2026,Chernyshev:2024pqy,spagnoli:prd2025}, though they can also be encoded into multi-qubit registers~\cite{spagnoli:prd2025}. Nevertheless, much can be learned from the two-flavor approximation, where each neutrino maps to a single qubit. To date, even the most sophisticated tensor network methods have been limited to simulating systems of only tens of neutrinos \cite{Cervia:2022pro}.

A central question for the efficient simulation of such systems is understanding where the quantum complexity resides throughout the evolution. Quantum entanglement has traditionally served as the primary diagnostic for ``quantumness" in many-body systems \cite{robin:prc2021, Roggero:2022hpy, Illa:2022zgu,cervia:prd2019,siwach:2023prd3}. For tensor-network methods such as matrix product states (MPS), the entanglement across any bipartition directly controls the bond dimension $d_\chi$ and hence the computational cost. In the context of collective neutrino oscillations, previous work established a direct connection between the growth of bipartite entanglement entropy and the emergence of spectral splits \cite{Patwardhan:2021rej,siwach:prd2025}. 

However, advances in quantum information science have made it clear that entanglement is not the only resource governing quantum complexity. A complementary measure is quantum non-stabilizerness, or ``magic”~\cite{Gottesman:arxiv1998,Aaronson:pra2004,Leone:prl2022}. An $N$-qubit stabilizer state is any state generated by the $N$-qubit Clifford group, comprising Hadamard, phase, and CNOT gates. The Gottesman–Knill theorem showed that these states—some of which are maximally entangled—can be simulated classically with complexity linear in the number of qubits \cite{Gottesman:arxiv1998}. Conversely, product states with zero entanglement can possess large magic, yet remain efficiently representable by tensor networks. These extreme cases suggest that entanglement and magic are orthogonal resources, occupying complementary regions of the quantum-complexity landscape. For generic many-body systems, however, the two quantities are far from independent: numerical studies across a range of models have revealed strong correlations between entanglement entropy and the stabilizer R\'enyi entropy (SRE), with the SRE often being less sensitive to bond-dimension truncation than the entanglement entropy itself~\cite{viscardi:spc2026}. This interplay implies that neither resource alone suffices to characterize the full computational difficulty of a many-body problem and motivates the search for a unified diagnostic~\cite{Masot-Llima:prl2024}.

A further open challenge is to connect these abstract resource measures to physical observables. Entanglement entropy admits well-known links to correlation functions and area law scaling, but magic lacks a comparably direct experimental or observational proxy. Recent studies have begun to probe magic in small neutrino systems, including three-flavor configurations \cite{Chernyshev:2024pqy, robin:arxiv2026}, yet the relationship between non-stabilizerness and the phenomenologically important spectral splits remains unexplored.


The purpose of this work is twofold. First, we aim to elucidate how quantum resources, that is, entanglement and magic, are distributed among the flavor modes of a dense neutrino gas as it evolves in time. Second, we investigate how the redistribution of these resources is reflected in physical observables, with particular emphasis on the emergence of spectral splits. We evaluate the von Neumann entropy, the non-local stabilizer R\'enyi entropy (non-local magic), and the MPS bond dimension for a variety of initial flavor compositions. We find that both entanglement entropy and non-local magic undergo a pronounced reorganization precisely at the spectral split frequency, and that the MPS bond dimension tracks this combined resource redistribution. These results establish a concrete link between the quantum resources that govern computational complexity and the observables relevant to supernova neutrino physics, offering guidance for the design of both classical algorithms and near-term quantum simulations of dense neutrino environments.

\section{Methods}
We consider a two-flavor  $N$-neutrino system in the regime where vacuum and two-body interactions dominate. The Hamiltonian in the mass basis is given by \cite{cervia:prd2019}
\begin{eqnarray}\label{eq:H}
    H=-\sum_\omega \omega J^z_\omega+\mu(t)\sum_{\omega,\omega', \omega\neq\omega'}\vec{J}_\omega\cdot\vec{J}_{\omega'}
\end{eqnarray}
where $\omega$ represents the vacuum oscillation frequency and $\vec{J}_{\omega}$ are the SU(2) neutrino isospin operators for a neutrino in $\omega$ frequency mode. $\mu(t)$ is the neutrino-neutrino interaction~\cite{Cervia:2022pro}. We transform the Hamiltonian into the flavor basis employing the PMNS matrix~\cite{Maki:1962mu} and perform the simulations in flavor basis under the time-dependent Schr\"odinger equation. The polarization vector of a neutrino with frequency $\omega$ is given by $\vec{P}(\omega)=2\bra{\Psi}\vec{J}_{\omega}\ket{\Psi}$ where $\ket{\Psi}$ is the total wave function.

We employ the tensor network methods to simulate the dynamics of neutrinos under the Hamiltonian given in Eq.~\eqref{eq:H}, especially the time-dependent variational principle(TDVP) with global subspace expansion(GSE)~\cite{Yang:prb2020} for MPS.

A measure of local entanglement within $|\Psi\rangle$ is the bipartite entanglement entropy given by the von Neumann entropy of $\rho_A$
\begin{equation}
    S(\rho_A)  = -\text{Tr}\left(\rho_A\ln \rho_A\right),
\end{equation}
where $\rho_A=\text{Tr}_B\;\rho$ is the reduced density matrix of partition $A$ obtained by tracing out system $B$.

Quantum magic is a measure of the non-stabilizerness of a general state $|\Psi\rangle$. An $N$-qubit stabilizer state $|\mathcal{S}\rangle$ is generated by the $N$-qubit Clifford set $\mathcal{C}$, whose elements are $H,\;S,$ and CNOT applied to all qubits and qubit pairs. The stabilizers of $|\mathcal{S}\rangle$ are Pauli strings $\sigma$ such that $\sigma|\mathcal{S}\rangle=|\mathcal{S}\rangle$. These stabilizers form a group $\mathcal{S}$ with $N$ generators. Stabilizer states are simulated with computational complexity linear in $N$ with the stabilizer tableau formalism \cite{Aaronson:pra2004,stim}.

It is well known that the Clifford set along with the non-Clifford $T$ gate form a universal quantum gate set, so naively a measure of magic of $|\Psi\rangle$ as proportional to the number of $T$ gates required to construct $|\Psi\rangle$. Nevertheless, this measure goes beyond quantum circuits to any quantum state. A more general measure of magic is the $n$-stabilizer R\'enyi entropy (SRE)
\begin{equation}
    M_\alpha(|\Psi\rangle) = \frac{1}{1-\alpha} \log\left(\sum_{P\in \mathcal{P}_N} \frac{\left|\langle\Psi|P|\Psi\rangle \right|^{2\alpha}}{2^N}\right),
\end{equation}
where $\mathcal{P}_N$ is the reduced $N$-qubit Pauli group with cardinality $4^N$. The SRE satisfies three conditions: (1) $M_\alpha(|\Psi\rangle)=0$ if and only if $|\Psi\rangle$ is a stabilizer state, $M_\alpha(C|\Psi\rangle)=M_\alpha(\Psi)$ for any Clifford gate $C$, (3) $M_\alpha(|\Psi\rangle\otimes|\phi\rangle) = M_\alpha(|\Psi\rangle) + M_\alpha(|\phi\rangle)$. With the complexity going like $4^N$, much work has been done to efficiently and accurately calculate this quantity~\cite{Leone:prl2022,Haug:quantum2023,Lami:prl2023,Haug:prb2023,Tarabunga:prxQ2023}.

Another issue with this measure is that it is basis dependent. To determine the irreducible complexity of our many-body problem, we will need a basis independent measure of magic. One such measure is the bipartite non-local magic \cite{Cao:prxquantum2024,Qian:pra2025}. For the case of a bipartite system, the non-local magic is given by
\begin{equation}
    M^{\text{(NL)}}_\alpha(|\Psi\rangle) = \min_{U_A\otimes U_B} M_\alpha(U_A\otimes U_B |\Psi\rangle),
\end{equation}
where $U_{A(B)}$ are unitary matrices that act exclusively on subsystem $A(B)$. Qualitatively, this measures the amount of magic that lives in the correlation between $A$ and $B$, where the minimization of $U_A\otimes U_B$ removes basis dependence. Cao et. al \cite{Cao:prxquantum2024} showed the bounds of the non-local magic are
\begin{equation}
\begin{split}
    \mathcal{F}(\rho_A)/8 &\leq M^{\text{(NL)}}_2(\rho) \leq M_2(\{\lambda_i\}),
\end{split}
\end{equation}
where the lower bound $\mathcal{F}(\rho_A)$ is the antiflatness
\begin{equation}
    \mathcal{F}(\rho_A) = \text{Tr}\;\rho_A^3 - \left(\text{Tr}\;\rho_A^2\right)^2,
\end{equation}
referring to the flatness of the entanglement spectrum $\{\lambda_i\}$. Robin and Savage \cite{Robin:2025ymq} have shown that $4\mathcal{F}$ in some cases can conform to an exact measure of non-local magic. We will use this as a tighter lower bound. The upper bound is defined by non-local stabilizer R\'enyi entropy (NL-SRE$_{2}$)in terms of the entanglement spectrum $\{\lambda_i\}$ of a bipartitioned $|\Psi\rangle$ as
\begin{equation}\label{eq:nl-magic}
\begin{split}
    M_2(\{\lambda_i\}) =& -\log\left( \sum_{i_1,i_2,i_3,i_4=0}^{r-1} \sqrt{\lambda_{i_1}\lambda_{i_2}\lambda_{i_3}\lambda_{i_4}}\right.\\
    &\qquad\qquad\times \sqrt{\lambda_{i_1\wedge i_2 \wedge i_3}\lambda_{i_1\wedge i_2 \wedge i_4}}\\
    &\qquad\qquad\left. \times \sqrt{\lambda_{i_1\wedge i_3 \wedge i_4}\lambda_{i_2\wedge i_3 \wedge i_4}} \right),
\end{split}
\end{equation}
where $r=2^{|A|}$ ($|A|$ is the size of partition $A$) is the dimension of the shared link between $A$ and $B$ and $\wedge$ is bitwise XOR addition. As the entanglement spectrum is independent of local unitaries acting exclusively on subsystems $A$ or $B$, $M_2(\{\lambda_i\})$ is independent of local unitaries $U_A$ and $U_B$, and is minimized when $\lambda_i\geq\lambda_{i+1}$ for $i\in\mathbb{Z}_r$.

\begin{figure}[h!]
    \centering
    \includegraphics[width=0.9\linewidth]{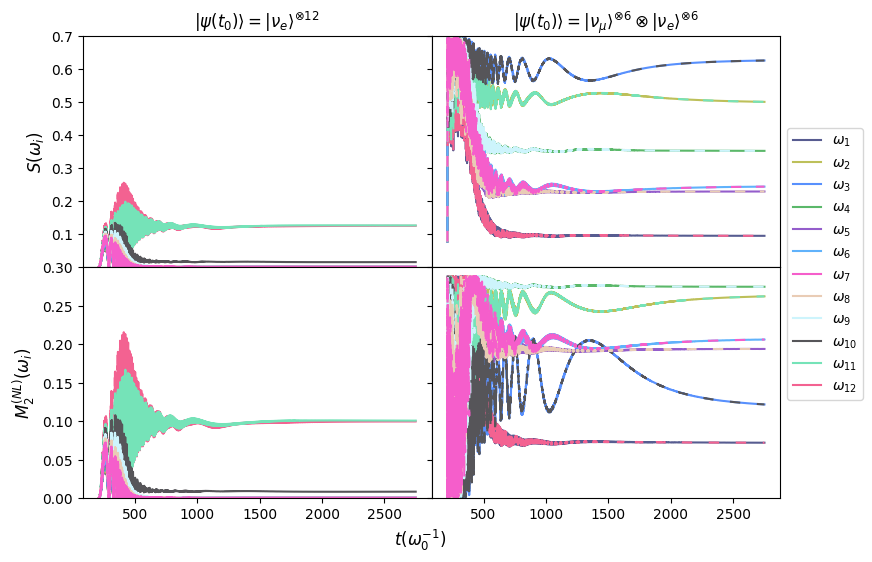}
    \caption{Dynamics of entanglement entropy(top panels) and non-local magic(bottom panels) for neutrinos with frequencies $\omega_i$ for initial states $|\Psi(t_0)\rangle=|\nu_e\rangle^{\otimes12}$(right panels) and $|\nu_\mu\rangle^{\otimes6}\otimes|\nu_e\rangle^{\otimes6}$ (left panels).}
    \label{fig:dynamics}
\end{figure}
\section{Results}
To analyze how the quantum resources are organized, we simulate an $N=12$ neutrino system. The entanglement entropy and non-local magic for an initial state with all neutrinos in electron flavor $|\Psi(t_0)\rangle=|\nu_e\rangle^{\otimes12}$ are shown in the left panels of Figure~\ref{fig:dynamics}. During the evolution, both the entanglement entropy and non-local magic develop strongly frequency-dependent structures. High frequency modes acquire the largest entanglement and non-local magic, while lower frequency modes remain comparatively weakly correlated. Neither resource reaches its maximal value, indicating that the system occupies an intermediate-complexity regime throughout the evolution. A qualitatively different structure emerges for mixed initial flavor configurations. For a more general initial state, $|\Psi(t_0)\rangle=|\nu_\mu\rangle^{\otimes6}\otimes|\nu_e\rangle^{\otimes6}$, the results are shown in the right panels of Figure~\ref{fig:dynamics}. Both entanglement entropy and non-local magic initially exhibit strong oscillatory behavior, with entanglement entropy reaching its maximal value of $\ln2$, and  the NL-SRE$_2$ oscillating between 0 and its maximum $\ln(4/3)$ (see Appendix \ref{app:constraint}). At late times, however, the resource distribution separates into distinct dynamical sectors, a subset of neutrino modes evolves toward low-entanglement, low-magic configurations, while other modes remain highly nonclassical.

\begin{figure}
    \centering
    \includegraphics[width=0.9\linewidth]{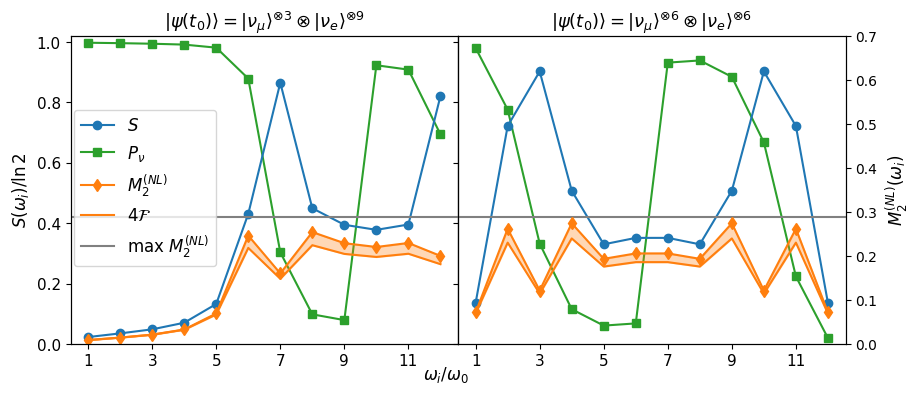}
    \caption{Asymptotic values of normalized entanglement entropy $S$, non-local magic $M^{(NL)}_2$, antiflatness $4\mathcal{F}$, and $P_\nu$ for a 12 neutrinos system with initial configuration $\ket{\Psi(t_0)}=\ket{\nu_\mu}^{\otimes3}\otimes\ket{\nu_e}^{\otimes9}$ (left) and $\ket{\Psi(t_0)}=\ket{\nu_\mu}^{\otimes6}\otimes\ket{\nu_e}^{\otimes6}$ (right).}
    \label{fig:asymp-vals}
\end{figure}

The most striking behavior occurs near the spectral split frequencies. To identify the spectral split~\footnote{Spectral splits occur when there is a sudden change in $P_{\nu_1}$. A stronger split occur when $|P_{\nu_1}(\omega_i)-P_{\nu_1}(\omega_{i\pm1})|\geq 0.5$ whereas for a weaker split $|P_{\nu_1}(\omega_i)-P_{\nu_1}(\omega_{i\pm1})|$ can be smaller.} locations, we analyze the probability $P_{\nu_1}$ of the first mass eigenstate in the asymptotic limit. For an initial configuration $|\Psi(t_0)\rangle=|\nu_\mu\rangle^{\otimes3}\otimes|\nu_e\rangle^{\otimes9}$, we observe a strong spectral split at $\omega_6\leftrightarrow\omega_7$ and a weaker one at $\omega_{11}\leftrightarrow\omega_{12}$ as shown in left panel of Figure~\ref{fig:asymp-vals}. The entanglement entropy at $\omega_7$ and $\omega_{12}$ is the maximum and significantly larger than the neighboring modes. Simultaneously, the non-local magic develops a local suppression at precisely the same frequencies. This behavior demonstrates that spectral splits are associated not simply with enhanced entanglement, but with a redistribution of quantum resources in flavor space. Away from the split frequencies, entanglement and non-local magic largely evolve in tandem. Near the splits, however, the two resources reorganize differently, revealing a restructuring of the underlying many-body complexity. The same pattern persists for more general initial flavor configurations $|\Psi(t_0)\rangle=|\nu_\mu\rangle^{\otimes6}\otimes|\nu_e\rangle^{\otimes6}$ as shown in the right panel of Figure~\ref{fig:asymp-vals}. Two strong spectral splits emerge at $\omega_2\leftrightarrow\omega_3$ and $\omega_{10}\leftrightarrow\omega_{11}$. In agreement with our previous observations, the neutrinos with frequencies $\omega_3$ and $\omega_{10}$ are maximally entangled and the non-local magic at these frequencies suppresses to a lower value as compared to the neighboring frequency neutrinos redistributing the resources at spectral splits. These findings are consistent for different initial states and system sizes (see Appendix \ref{app:extend-results}). Therefore, at the split, the state becomes locally complex (high entanglement) but structurally simpler in its non-local correlations (low magic), suggesting the spectral split acts as a ``complexity phase transition."

\begin{figure}
    \centering
    \includegraphics[width=0.6\linewidth]{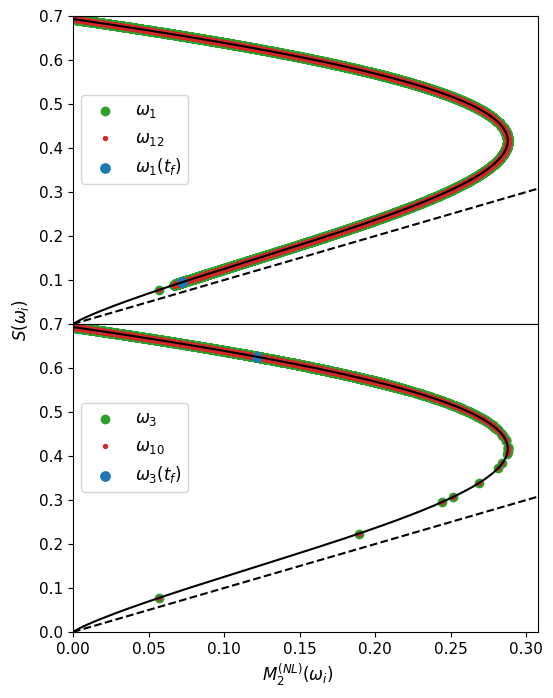}
    \caption{Dynamics of non-local magic and entanglement entropy of neutrinos with frequency modes $\omega_1(\omega_{12})$ and $\omega_3(\omega_{10})$ in phase space for the initial configuration $\ket{\Psi(t_0)}=\ket{\nu_\mu}^{\otimes6}\otimes\ket{\nu_e}^{\otimes6}$. The solid line comes from the constraint on the entanglement spectra that $\lambda_0+\lambda_1=1$. The dashed line is at a $45^{\circ}$ angle to emphasize the position of arc in phase space. The final position of the entropy and magic are denoted by the blue dots on the arc.}
    \label{fig:dynamics-vals}
\end{figure}

To further investigate the resource dynamics, we analyze the evolution in the entanglement-magic phase space. The resource dynamics for the initial configuration $|\Psi(t_0)\rangle=|\nu_\mu\rangle^{\otimes6}\otimes|\nu_e\rangle^{\otimes6}$ and neutrinos with frequencies $\omega_1(\omega_{12})$ and $\omega_3(\omega_{10})$ are shown in Figure~\ref{fig:dynamics-vals}. Individual neutrino modes follow constrained trajectories forming characteristic arcs and the symmetry between two halves of the system is persistent. By measuring the absolute difference between the pairs of neutrinos with same dynamic for both quantities over the course of the evolution, pairs agree to within $10^{-5}$, confirming the symmetry is exact within numerical precision. The arc emerges from the fact that the entanglement spectrum is normalized. Modes not associated with spectral splits, that is, $\omega_1(\omega_{12})$ explore broad regions of the phase space spanning low- and high-resource sectors. By contrast, modes near the spectral split frequencies, that is, $\omega_3(\omega_{10})$ remain confined predominantly to highly entangled regions. Notably, the system never accesses the simultaneous maximum-entanglement and maximum-magic regime. Instead, the evolution preferentially explores sectors in which increases in entanglement are accompanied by suppression or redistribution of non-local magic. This observation further supports the interpretation of spectral splits as signatures of nontrivial resource reorganization.


\begin{figure}
    \centering
    \includegraphics[width=0.6\linewidth]{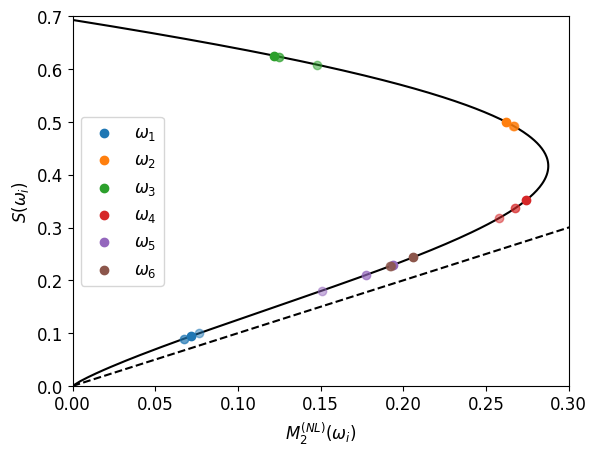}
    \caption{Dependence of non-local magic and entanglement entropy on bond dimension in the asymptotic limit for neutrinos with frequency modes $\omega_1$ through $\omega_6$ for the initial configuration $\ket{\Psi(t_0)}=\ket{\nu_{\mu}}^{\otimes6}\otimes\ket{\nu_e}^{\otimes6}$. Opacity is proportional to bond dimension, corresponding to $\max d=64,48,$ and $36$ respectively.}
    \label{fig:EEvsSRE}
\end{figure}

Finally, we examine the relationship between bond dimension, entanglement entropy, and non-local magic in Figure \ref{fig:EEvsSRE}. In the high entanglement regime ($S\gtrsim0.4$), both resources anti-correlate, for example, if entanglement entropy increases(decreases) non-local magic decreases(increases) with a reduction in the bond dimension. However, in the low entanglement regime($S\lesssim0.4$), both resources behave in tandem, that is, if entanglement entropy increases, non-local magic also increases. 
For example, for neutrinos with frequency modes $\omega_2$ and $\omega_3$, which lie in the high entanglement regime, entanglement entropy decreases but non-local magic increases with a reduction in the bond dimension. For all other neutrinos, which lie in the lower entanglement regime, entanglement entropy and non-local magic both either decrease or increase with a decrease in the bond dimension. Therefore, the bond dimension consistently tracks the overall restructuring of the quantum resource landscape. In particular, the geometric structure of the entanglement-magic phase space remains robust under bond dimension truncation. This indicates that the MPS bond dimension provides an operational probe of the underlying many-body resource structure.

\section{Conclusion}
In summary, we have demonstrated that spectral splits in collective neutrino oscillations are characterized by a distinctive quantum-resource signature, that is, maximal bipartite entanglement coinciding with locally minimal non-local magic. This anticorrelation reveals that splits emerge through structured resource redistribution rather than monotonic complexity growth. The constrained arc dynamics in entanglement-magic phase space, governed by entanglement-spectrum normalization, provide a geometric picture of the accessible resource landscape. These findings create a concrete bridge between an astrophysical phenomenon and the underlying quantum information resources, offering a clear roadmap for the development of more efficient classical tensor network methods and near‑term quantum simulators of dense neutrino gases. For example, classical tensor network methods should target the split frequency where bond-dimension requirements peak, while quantum circuits may exploit the reduced magic at splits to minimize non-Clifford gate overhead. Extensions to three flavors, larger systems, and connections to stabilizer tensor networks \cite{Masot-Llima:prl2024} remain important directions.



\begin{acknowledgments}
Michael Hite (MH) thanks the organizers and participants of the iQuS workshop From String Dynamics to Event Generators with Quantum Simulation for useful discussions on quantum complexity. MH also thanks ChunJun Cao for clarifying discussions on non local magic measures. This material is based upon work supported in part by the U.S. Department of Energy, Office of Science, Office of Nuclear Physics, under the FRIB Theory Alliance award DE-SC0013617.
\end{acknowledgments}

\appendix
\section{Matrix product state}
In MPS, an $N$-body quantum state $|\psi\rangle$ is expressed in terms of a tensor network with $N$ tensors as
\begin{equation}
    \psi_{i_1\dots i_N} = \psi^{(1)}_{i_1\chi_1} \psi^{(2)}_{\chi_1i_2\chi_2}\dots \psi^{(N-1)}_{\chi_{N-2}i_{N-1}\chi_{N-1}} \psi^{(N)}_{\chi_{N-1}i_N},
\end{equation}
where $i_1,\dots,i_N$ are the physical indices of dimension $d_i=2$ (qubits) and $\chi_1,\dots,\chi_{N-1}$ are bond indices whose dimensions $d_\chi$ are dependent upon the entanglement of $|\psi\rangle$. An $N$-qubit product state will have $\max d_\chi=1$, while the maximally entangled state will have $\max d_\chi=2^{N/2}$.

\section{Constraint in Quantum Complexity Phase Space}\label{app:constraint}
\begin{figure}[h]
    \centering
    \includegraphics[width=0.6\linewidth]{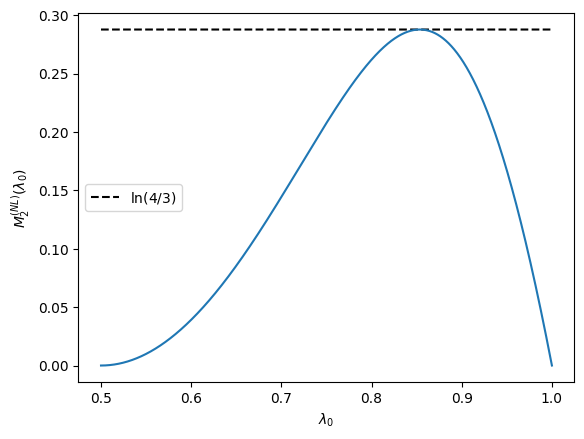}
    \caption{Equation \ref{eq:NLSRE2} in the interval $1/2\leq\lambda_0\leq1$. It reaches its maximal value of $\ln(4/3)$ at $\lambda_0=\sqrt{2}/4+1/2$.}
    \label{fig:NLSRE2}
\end{figure}

Consider the bipartitions $A$ and $B$ of an $N$ qubit system of sizes $|A|=1$ and $|B|=N-1$. The entanglement spectrum (singular values of the reduced density matrix) contains two values $\lambda_0,\lambda_1$ such that $\lambda_0+\lambda_1=1$. The condition in the calculation of non-local 2-stabilizer R\'enyi entropy (NL-SRE$_2$) that $\lambda_0\geq\lambda_1$ means $1/2\leq \lambda_0 \leq 1$ and $0\leq \lambda_1 \leq 1/2$. The NL-SRE$_2$ in terms of $\lambda_0$ is given by
\begin{equation}
\begin{split}\label{eq:NLSRE2}
    M^{(NL)}_2(\lambda_0) &= -\ln\left(\lambda_0^4+\lambda_1^4 + 14\lambda_0^2\lambda_1^2\right)\\
    &= -\ln\left(\lambda_0^4+(1-\lambda_0)^4 + 14\lambda_0^2(1-\lambda_0)^2\right).
\end{split}
\end{equation}
The function is minimized at $\lambda_0=\sqrt{2}/4+1/2$ or $M_2^{(NL)}=\ln(4/3)$ (Figure \ref{fig:NLSRE2}). Similarly, the bipartite entanglement entropy is
\begin{equation}
    S(\lambda_0) = -\lambda_0\ln\lambda_0 - \lambda_1\ln\lambda_1= -\lambda_0\ln\lambda_0 - (1-\lambda_0)\ln(1-\lambda_0),
\end{equation}
which reaches it maximum value of $\log2$ at $\lambda_0=1/2$. The arc in Figures 3 and 4 of the main article is then $(M_2^{(NL)}(\lambda_0), S(\lambda_0))$ for $1/2\leq \lambda_0 \leq 1$.

\section{Extended Results}\label{app:extend-results}
\begin{figure}[h]
    \centering
    \includegraphics[width=0.9\columnwidth]{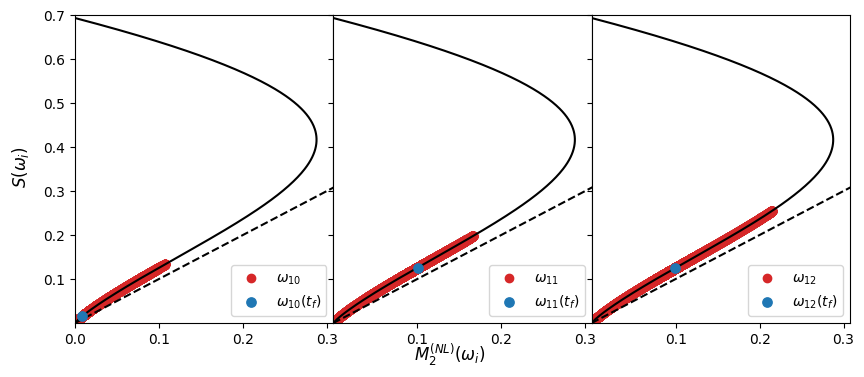}
    \caption{Dynamics of modes $\omega_{10}$, $\omega_{11}$, and $\omega_{12}$ for initial state $|\psi(t_0)\rangle=|\nu_e\rangle^{\otimes12}$. The blue dot corresponds to the value in the asymptotic limit ($t=2750.64$). The dashed line separates the entanglement/magic dominant regimes.}
    \label{fig:EE-vs-SRE-e12}
\end{figure}

Here we include results for more initial states and system sizes from the main article. We first consider the dynamics of the NL-SRE$_2$ and bipartite entanglement entropy in the all electron flavor initial state for $N=12$. We know from Figure \ref{fig:dynamics} that all frequencies $\omega_i$ evolve in a low magic/entanglement regime. Considering that only frequencies 10,11,and 12 end in a state of nonzero complexity, we look at their evolution in quantum complexity phase space in Figure \ref{fig:EE-vs-SRE-e12}. In this regime, NL-SRE$_2$ and entanglement entropy are proportional to one another, so an increase in one resource will always result in the increase of another. Figure \ref{fig:bonddim-N10} shows the change in complexity in the asymptotic limit through reducing the bond dimension for the initial configuration $|\psi(t_0)\rangle=|\nu_\mu\rangle^{\otimes5}\otimes|\nu_e\rangle^{\otimes5}$. Modes $\omega_1$ and $\omega_2$ increase in magic and entanglement as bond dimension increases, while mode $\omega_4$ decreases in magic and entanglement. Modes $\omega_2$ and $\omega_3$, lying in the high entanglement regime, move in opposite direction along the arc. The relationship between these resources and bond dimension is non trivial, as it seems to not depend upon which region of phase space they're in.

\begin{figure}[h]
\centering
    \includegraphics[width=0.8\columnwidth]{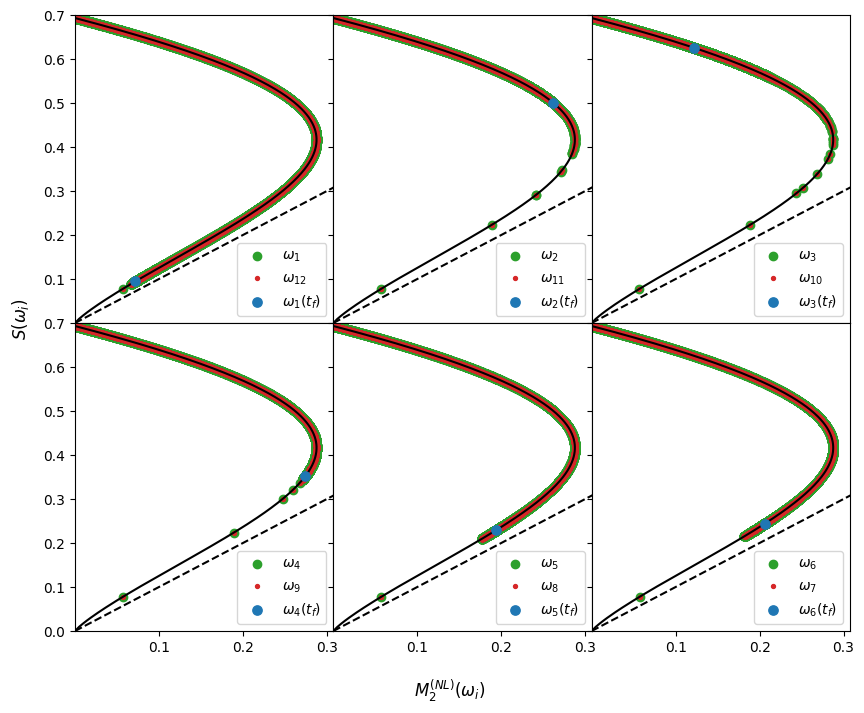}
    \caption{Dynamics of neutrino pairs in quantum complexity phase space for the initial state $|\psi(t_0)\rangle=|\nu_\mu\rangle^{\otimes6}\otimes|\nu_e\rangle^{\otimes6}$.}
    \label{fig:EE-vs-SRE-mu6-e6}
\end{figure}

\begin{figure}[h]
    \centering
    \includegraphics[width=0.6\linewidth]{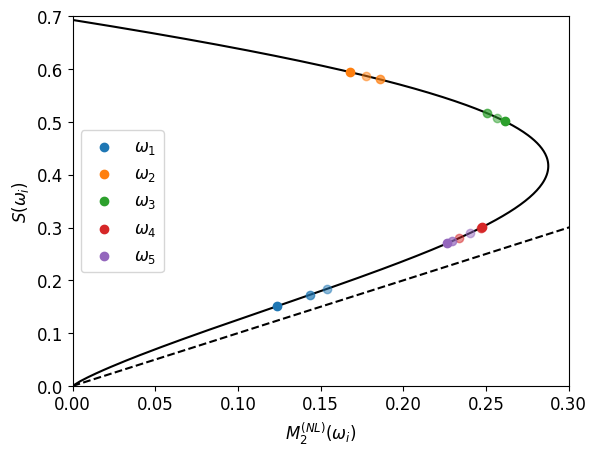}
    \caption{Dependence on bond dimension for frequency modes 1-5 in the asymptotic limit for the initial state $|\psi(t_0)\rangle=|\nu_\mu\rangle^{\otimes5}\otimes|\nu_e\rangle^{\otimes5}$. Opacity is proportional to bond dimension (max $d_\chi=32,28,26$).}
    \label{fig:bonddim-N10}
\end{figure}

Figure \ref{fig:EE-vs-SRE-mu6-e6} shows the evolution of all the frequency pairs for the initial state $|\psi(t_0)\rangle=|\nu_\mu\rangle^{\otimes6}\otimes|\nu_e\rangle^{\otimes6}$. We see that all pairs except $\omega_1(\omega_{12})$ live in the highly entangled regime nearly the entire evolution. The few points we see in the low entanglement regime are from the first few time steps as the states make their way up the arc when the neutrino-neutrino interaction term is dominant. 

\begin{figure}[h]
    \centering
    \includegraphics[width=1.0\columnwidth]{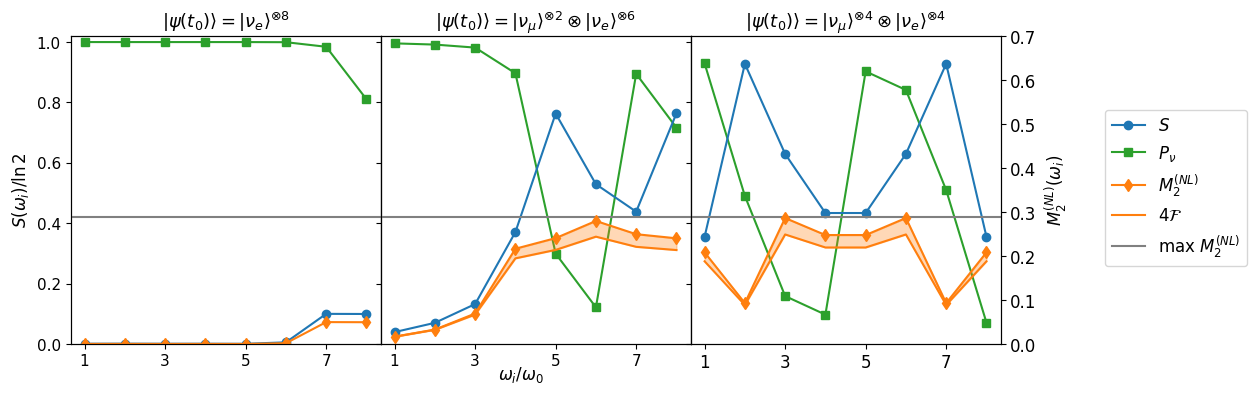}
    \caption{Asymptotic values of survival probability $P_\nu$, entanglement entropy $S$, non-local magic $M_2^{(NL)}$, and antiflatness for an $N=8$-site system with three different initial configurations.}
    \label{fig:asymp-vals-N8}
\end{figure}

\begin{figure}[h]
    \centering
    \includegraphics[width=1.0\columnwidth]{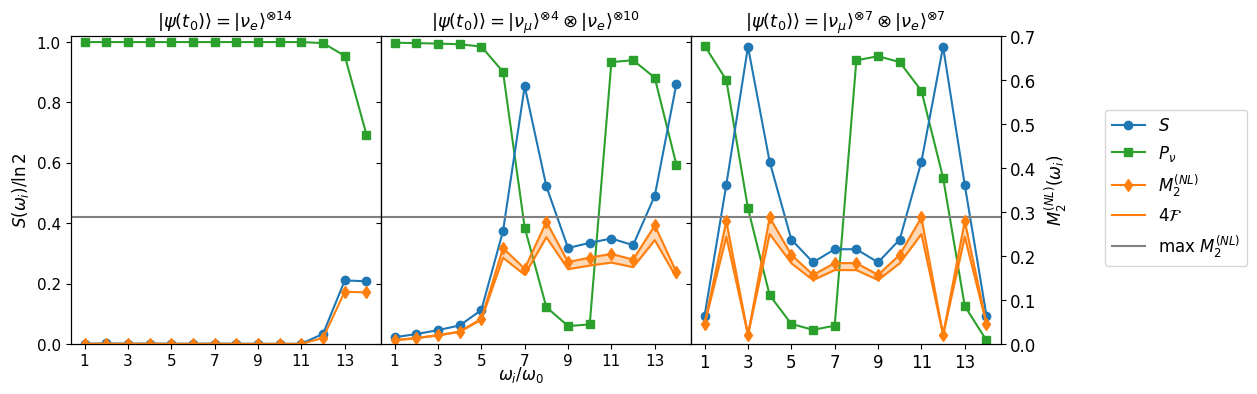}
    \caption{Asymptotic values of survival probability $P_\nu$, entanglement entropy $S$, non-local magic $M_2^{(NL)}$, and antiflatness for an $N=14$-site system with three different initial configurations.}
    \label{fig:asymp-vals-N14}
\end{figure}

We show asymptotic behavior for three different initial configuration for $N=8$ in Figure \ref{fig:asymp-vals-N8} and $N=14$ in Figure \ref{fig:asymp-vals-N14}. In the all same flavor initial states (left panels), there is no spectral split with a decrease in survival probability corresponding to an increase in complexity. For the initial configuration $|\psi(t_0)\rangle=|\nu_e\rangle^{\otimes2}\otimes|\nu_\mu\rangle^{\otimes6}$ (middle panel), the spectral split behavior of the NL-SRE$_2$ at $\omega_5$ is not a clear minimum like the other cases, but still indicative of a redistribution of resources. This may be due to the organization of the resources after the split, as the entanglement in $\omega_6$, $\omega_7$, and $\omega_8$ are quite different than the other cases. For $N=14$ multi-flavor initial states (middle and right panels), the value of NL-SRE$_2$ is a pronounced minima at the spectral split.

\clearpage
\bibliography{apssamp}

\end{document}